\documentstyle[aps,prl,psfig]{revtex}
\def\be{\begin{equation}}           
\def\ee{\end{equation}}
\def\bt{\begin{table}}           
\def\et{\end{table}}
\def\lsim{\lower0.5ex\hbox{$\; \buildrel < \over \sim \;$}}

\def\r{\rho}
\def\om{\omega}
\def\f{\frac}
\def\al{\alpha}
\def\b{\beta}
\def\pr{\prime}
\def\g{\gamma}
\def\d{\delta}
\def\ep{\epsilon}

\begin{document}

\draft

\title{Non-linear Regge spectrum fits to experimentally up-dated meson states}
\author{Jishnu Dey}
\address{Abdus Salam ICTP, Trieste, Italy and Azad
Physics Centre, Maulana Azad College, Calcutta 700 013, India}
\author{Mira Dey}
\address{Department of Physics, Presidency College, Calcutta 700 073, India,
and an associate member, Abdus Salam ICTP, Trieste, Italy}
\author{P. Leal Ferreira}
\address{IFT-UNESP, 145 Rua Pamplona,  Sao Paulo 01405-900, Brasil}
\author{L. Tomio}
\address{IFT-UNESP, 145 Rua Pamplona,  Sao Paulo 01405-900, Brasil}

\date{\today}
\maketitle

\begin{abstract}

The exciting feature of recent times is the inflow of new experimental data. Fits to 
the $\pi$, $\r$ and $f$ meson resonances found by the VES collaboration, along with 
those of the PDG 1998 compilation are reported using a non-linear model for Regge 
trajectories. There are only five parameters including the pion mass and the states 
come out as Regge -type excitations of the pion. Total number of mesons fitted are 23 
and the fit is good to $\sim$ 3\% for 18 of them and  $\sim$ 9\% for others.    

\end{abstract}

We have suggested the use of a modified Regge trajectory for heavy as well as light 
mesons and in the case of the upsilon radial excitations the non-linearity is most 
prominent \cite{ddft}. The trajectories are based on the use of deformed Poincar\'e 
algebra for the internal degrees of freedom of the hadron. In fact it was shown 
recently \cite{dbrr} that the kinetic excitation of the hadron must obey the usual 
Einstein relation in order that the thermodynamic functions can be defined for the 
system, and that the limiting temperature predicted by the fit is larger than the 
older Hagedorn estimate.   

The square of masses of the resonances is given in the model by
\be
E^2 = Sinh^{-1}[sinh^2(\f{m\ep}{2}) +(\f{\ep^2}{4})(\f{L}{\al^{\pr}} + \f{L}{\b^{\pr}} 
+ \f{S}{\g^{\pr}} + \f{J}{\d^{\pr}})]
\label{eq:1}
\ee
The parameters we choose are $\al^\pr = 0.72$, $ \b^\pr = 0.62$, $\g^\pr = 2.1$ and 
$\d^\pr = 9.2 $, all in $GeV^{-2}$. The parameter $\ep$ is taken to be 0.912 $GeV^{-1}
$ as found in \cite{ddft} from the upsilon fit. The states are given in the following 
tables. 

Table 1 gives the pion states. Note the state $\pi(1740)$ given by \cite{valeri}, 
revised from the PDG value $1800$ fits exceedingly well with our model. These states, 
revised by \cite{valeri} and \cite{db} was the motivation for this addendum and are 
shown by us with a superscript asterisk, in the Tables.   

We have not tried to fit the $\r$ and $a$ separately from the $\om$ and the $f$ 
mesons, since they are roughly degenerate. This could be done easily by choosing the
 $\al^\pr \; , \; \b^\pr\; , \;\g^\pr $ and $\d^\pr  $ separately for the two sets but 
is not otherwise very meaningful. There are very few omega meson states (4) and the $f$
mesons are excitations of both $\om$ and $\eta$. The new experiments do not report 
$\eta$ mesons and they are not considered in this paper. But all the mesons as well as 
baryons can be fitted in the model \cite{dbrr}. The interesting point is the inflow 
of the more recent data.    
   
The Table 2 shows the fits to the $\r\,- \,\om$ system. Note that the new $a_1(1800)$
state reported by \cite{valeri} with an experimental error of 50 MeV and width 230 
MeV, fits very well into our scheme as a nodal excitation, n =1, in the L = 1 channel.

\bt
\caption{The $\pi$ meson states in $MeV$.}
\vskip 1cm
\begin{center}
\begin{tabular}{|c|c|c|c|c|c|}
\hline
Experiment&Our fit&Comment & Experiment&Our fit&Comment\\
\hline
$\pi(139)$ &  139 & fitted &$b(1235)$& 1200 & error 3\% \\
$\pi(1300)$ & 1266& error 3\% & $\pi(1740)^*$& 1743& error 0.2\%\\
$\pi2(1670)$ & 1658& error 0.8\% & $\pi(2100)$& 2067& error 2\%\\
\hline 
\end{tabular}
\end{center}
\et 
\bt
\caption{The $\r$, $a$ and $\om$, $f$ mesons in $MeV$.}
\vskip 1cm
\begin{center}
\begin{tabular}{|c|c|c|c|c|c|}
\hline
Experiment&Our fit&Comment & Experiment&Our fit&Comment\\
\hline
$\r(770)$ &  776 & 1\% &$\om(782)$& 776 & error 1\% \\
$a_1(1260)$ & 1374& error 9\% & $f_1(1285)$ &1374 & error 9\%\\
$a_2(1360)$ & 1428& error 5\%&$f_2(1270)$ & 1410& error 10\%\\
$\r _3(1690)$ & 1799& error 8\%&$\om_3(1670)$ & 1799& error 9\%\\
$\r _3(2180)^*$ & 2108& error 4\%&$\r(1450)$&1465&error 1\%\\
$\r _3(2300)$ &2330& error 2\%&$a_1(1800)^*$&1813 &error 0.5\%\\
$a_4(2040)$ & 2079& error 2\%&$f_4(2050)$ & 2080& error 1.5\%\\
$a_6(2450)$ & 2464& error 0.5\%&$f_6(2510)$ & 2464& error 2\%\\
$\r _5(2350)$ & 2293& error 3\%&$\r _3(2250)$ & 2264& error 0.5\%\\
\hline 
\end{tabular}
\end{center}
\et

\end{document}